\documentclass[a4paper,11pt]{article}
\pdfoutput=1 
\usepackage{jheppub} 
\usepackage[T1]{fontenc} 
\usepackage{amsmath,amssymb,epsfig,txfonts}

\allowdisplaybreaks

\definecolor{nicered}{rgb}{0.7,0.1,0.1}
\definecolor{nicegreen}{rgb}{0.1,0.5,0.1}
\definecolor{niceblue}{rgb}{0.0,0.1,0.7}
\hypersetup{colorlinks,citecolor=nicegreen,linkcolor=nicered,urlcolor=niceblue}

\title{Collider constraints on light pseudoscalars}

\author[1,2]{Ulrich Haisch,}

\author[3,4]{Jernej F. Kamenik,}

\author[1]{Augustinas Malinauskas,}

\author[5]{and Michael Spira}

\affiliation[1]{Rudolf Peierls Centre for Theoretical Physics,
   University of Oxford, OX1 3NP Oxford, United Kingdom}
    
\affiliation[2]{CERN, Theoretical Physics Department,  CH-1211 Geneva 23, Switzerland}

\affiliation[3]{Jo\v zef Stefan Institute, Jamova 39, 1000 Ljubljana, Slovenia}

\affiliation[4]{Faculty of Mathematics and Physics, University of Ljubljana, Jadranska 19, 1000 Ljubljana, Slovenia}

\affiliation[5]{Paul Scherrer Institut, CH-5232 Villigen PSI, Switzerland}

\emailAdd{ulrich.haisch@physics.ox.ac.uk}
\emailAdd{jernej.kamenik@cern.ch}
\emailAdd{augustinas.malinauskas@physics.ox.ac.uk}
\emailAdd{michael.spira@psi.ch}

\abstract{

\noindent We investigate   the bounds on light pseudoscalars that arise from a variety of  collider searches. Special attention is thereby devoted to the mass regions $[3, 5] \, {\rm GeV}$ and $[9,11] \, {\rm GeV}$, in which a meaningful theoretical description has to include estimates of non-perturbative effects such as the mixing of the pseudoscalar with QCD bound states. A~compendium of formulas that allows to deal with the relevant corrections  is provided. It  should prove useful for the interpretation of future LHC  searches for light CP-odd spin-0 states.}

\preprint{CERN-TH-2018-025, PSI-PR-18-03}

\begin{document} 

\maketitle

\section{Introduction}
\label{sec:introduction}

The most significant achievement of the LHC Run-I  physics programme has been the discovery of a new spin-0 resonance~($h$) with a mass of  $125 \, {\rm GeV}$ and with properties consistent with that of the standard model (SM) Higgs boson~\cite{Aad:2012tfa,Chatrchyan:2012xdj,Khachatryan:2016vau}. Besides precision measurements of processes involving a $h$, the LHC Higgs physics programme however also includes a wide spectrum of searches for additional Higgses (a~summary of LHC Run-I results can be found in~\cite{CMS:2016qbe} for instance). Such states  are predicted in many SM extensions such as supersymmetry  or models where the Higgs is realised as a pseudo Nambu-Goldstone boson~(PNGB) of a new approximate global symmetry. 

In fact, if the extended electroweak~(EW) symmetry breaking sector contains a PNGB, this state can be significantly lighter than the other spin-0 particles. A well-known example of a model that includes a light pseudoscalar~($a$) is provided by the next-to-minimal supersymmetric SM~(NMSSM) where this state can arise as a result of an approximate global $U(1)_R$ symmetry~\cite{Dobrescu:2000yn}. Since in this case the amount of symmetry breaking turns out to be proportional to soft breaking trilinear terms,  the  mass of the $a$ can  naturally be less than half of the SM Higgs mass, if the trilinear terms are  dialled to take values in the GeV range. Non-supersymmetric theories that can feature a light pseudoscalar are, to just name a few,   simplified models where  a complex singlet scalar is coupled to the Higgs potential of the SM or the  two-Higgs doublet model~(2HDM), Little Higgs models and hidden valley scenarios (see~\cite{Curtin:2013fra} and references therein for details).

Irrespectively of the precise ultraviolet~(UV) realisation, a light pseudoscalar  can lead to distinctive collider signatures. The most obvious consequence are exotic   decays of the SM Higgs, namely $h \to aa$ for $m_a < m_h/2$~\cite{Dobrescu:2000jt,Dermisek:2005ar} and $h \to aZ$ for $m_a < m_h - m_Z$~\cite{Curtin:2013fra,Christensen:2013dra}.  Another feature that  can have important phenomenological implications is that  in the presence of the heavy-quark transition $a \to b \bar b$ ($a \to c \bar c$) the pseudoscalar $a$ can mix with bottomonium (charmonium) bound states with matching quantum numbers~\cite{Drees:1989du,Domingo:2008rr,Domingo:2010am,Domingo:2011rn,Baumgart:2012pj,Haisch:2016hzu,Domingo:2016yih}.

LHC searches for $h \to aa$ have been performed in the $4 \mu$~\cite{Khachatryan:2015wka,CMS:2016tgd}, $4 \tau$~\cite{Khachatryan:2015nba,Khachatryan:2017mnf}, $2 \mu 2\tau$~\cite{Khachatryan:2017mnf,CMS-PAS-HIG-17-029}, $2 \mu 2b$~\cite{Khachatryan:2017mnf} and $2 \tau 2 b$ final states~\cite{CMS-PAS-HIG-17-024}.  The obtained results have been used to set upper bounds on the $h \to aa$ branching ratio in 2HDMs with an extra complex singlet~(2HDM+S) for pseudoscalar masses in the range of $[1, 62.5] \, {\rm GeV}$. The analyses~\cite{Khachatryan:2015wka,Khachatryan:2015nba,Khachatryan:2017mnf,CMS-PAS-HIG-17-029,CMS-PAS-HIG-17-024} however all exclude $m_a$ values in the regions $[3,5] \, {\rm GeV}$ and~$[9,11] \, {\rm GeV}$ for which $a \hspace{0.25mm}$--$\hspace{0.5mm} \eta_c$ and $a \hspace{0.25mm}$--$\hspace{0.5mm} \eta_b$ mixing effects as well as open flavour decays to $D$ and~$B_{(s)}$ meson pairs can be potentially important. 

The main goal of this work is to extend the latter results to the $c \bar c$ and $b \bar b$ threshold regions by including effects that  cannot be properly described in the partonic picture. In order to highlight the complementarity of  different search strategies for a light $a$, we also compare our improved limits  to other  bounds on the~2HDM+S parameter space that derive from the LHC searches for $h \to Z_d Z \to 4 \ell$~\cite{Aad:2015sva}, $h \to Z_d Z \to 2\mu 2 \ell$~\cite{Aaboud:2018fvk}, $pp \to a \to \mu^+ \mu^-$~\cite{ATLAS:2011cea,Chatrchyan:2012am}, $pp \to a b \bar b$ followed by $a \to \tau^+ \tau^-$~\cite{Khachatryan:2015baw} or  $a \to \mu^+ \mu^-$~\cite{Sirunyan:2017uvf}, $pp \to a \to \gamma \gamma$~\cite{CMS-PAS-HIG-17-013,Mariotti:2017vtv}, $pp \to a \to \tau^+ \tau^-$~\cite{CMS-PAS-HIG-16-037}, from the BaBar analyses of  radiative $\Upsilon$ decays~\cite{Lees:2011wb,Lees:2012iw,Lees:2012te} and from the LHCb measurements of  the production of $\Upsilon$ mesons~\cite{Haisch:2016hzu,Aaij:2015awa} as well as the inclusive dimuon cross section~\cite{Ilten:2016tkc,Aaij:2017rft}. 

This article is organised as follows. In Section~\ref{sec:generalities} we briefly recall  the structure of the 2HDM+S scenarios. Our recast of the results~\cite{Khachatryan:2015wka,Khachatryan:2015nba,Khachatryan:2017mnf,CMS-PAS-HIG-17-029,CMS-PAS-HIG-17-024} is presented in Section~\ref{sec:results}, where we also derive the constraints on the 2HDM+S parameter space that follow from the measurements and prosposals~\cite{Haisch:2016hzu,Aad:2015sva,Aaboud:2018fvk,ATLAS:2011cea,Chatrchyan:2012am,Khachatryan:2015baw,Sirunyan:2017uvf,CMS-PAS-HIG-17-013,Mariotti:2017vtv,CMS-PAS-HIG-16-037,Lees:2011wb,Lees:2012iw,Lees:2012te,Aaij:2015awa,Ilten:2016tkc,Aaij:2017rft}. We conclude in~Section~\ref{sec:conclusions}. The formulas necessary to calculate the partial decay widths of the  pseudoscalar~$a$ are collected  in~Appendix~\ref{app:widths}, while Appendix~\ref{app:mixing} contains a concise discussion of the mixing formalism and of open flavour decays that are relevant in the vicinity of the $b \bar b$ and $c \bar c$ thresholds. 

\section{Theoretical framework}
\label{sec:generalities}

In the following section we will interpret various searches for light pseudoscalars in the context of 2HDM+S scenarios. In this class of models a complex scalar singlet~$S$ is added to the 2HDM Higgs potential~(see~e.g.~\cite{Gunion:1989we,Branco:2011iw} for 2HDM reviews). The field $S$  couples only to the two Higgs doublets~$H_{1,2}$ but has no direct Yukawa couplings, acquiring all of its couplings to SM fermions through its mixing with the Higgs doublets. A light pseudoscalar $a$ can arise in such a setup from the admixture of the 2HDM pseudoscalar $A$ and the imaginary part of the complex singlet $S$. The corresponding mixing angle will be denoted by $\theta$, and defined such  that for $\theta \to 0$ the mass eigenstate $a$ becomes exactly singlet-like. 

In order to eliminate phenomenologically dangerous tree-level flavour-changing neutral currents~(FCNCs) the Yukawa interactions that involve the Higgs fields $H_{1,2}$  have to satisfy the  natural flavour conservation hypothesis~\cite{Glashow:1976nt,Paschos:1976ay}.  Depending on which fermions couple to which doublet, one can divide the resulting 2HDMs into four different types. In~all four cases the Yukawa couplings between the pseudoscalar $a$ and the SM fermions take the generic form 
\begin{equation} \label{eq:La}
{\cal L} \supset - \sum_f \frac{y_f}{\sqrt{2}} \, i \hspace{0.25mm} \xi_f^{\rm M} \, \bar f \gamma_5 f \hspace{0.25mm} a \,.
\end{equation}
Here $y_f = \sqrt{2} m_f/v$ denote the SM Yukawa couplings and $v \simeq 246 \, {\rm GeV}$ is the EW vacuum expectation value. The parameters  $\xi_f^{\rm M}$ encode the dependence on the 2HDM Yukawa sector and the factors relevant for the further discussion are given in Table~\ref{tab:xifM}. In this table the shorthand notations $s_\theta = \sin \theta$ and $t_\beta = \tan \beta$ have been used.  Similar abbreviations will also be used in what follows. 

In the presence of~(\ref{eq:La}) the CP-odd scalar $a$ can decay into fermions at tree level and into gluons, photons and EW gauge bosons at loop level. The expressions for the partial decay widths $\Gamma  \hspace{0.25mm}  ( a \to XX )$ that we employ in our study are given in Appendix~\ref{app:widths}. Since in this work we will assume that the  $a$ is lighter than the $W$, $Z$, $h$ and the other 2HDM Higgs mass eigenstates $H$, $A$, $H^\pm$, decays of the $a$ into the latter states are kinematically forbidden. 

If the $a$ is sufficiently light, exotic decays of the SM Higgs into the two final states $aZ$ and $aa$ are however possible.  The partial decay width $\Gamma  \hspace{0.25mm}  ( h \to aZ  )$  is in  2HDM+S scenarios entirely fixed by the 2HDM parameters $\alpha, \beta$ and the mixing angle~$\theta$.  Explicitly, one has at tree level 
\begin{equation}
\Gamma  \hspace{0.25mm}  ( h \to aZ ) = \frac{g_{haZ}^2}{16 \pi} \hspace{0.25mm} \frac{m_h^3}{v^2} \hspace{0.25mm} \lambda^3 \hspace{-0.5mm} \left (m_h^2,m_a^2,m_Z^2 \right ) \,,
\end{equation}
with 
\begin{equation} \label{eq:ghaZ}
g_{haZ} = c_{\beta - \alpha}  \hspace{0.5mm} s_\theta \,,
\end{equation}
and
\begin{equation} \label{eq:lambdaxyz}
\lambda \left (x, y, z \right ) = \sqrt{1  - \frac{2 \left (y + z \right )}{x}  + \frac{(y - z)^2}{x^2} } \,.
\end{equation}
Notice that in the exact alignment/decoupling limit, i.e.~$\alpha =  \beta - \pi/2$, in which the lighter CP-even spin-0 state~$h$  of the 2HDM becomes fully SM-like, the coupling $g_{haZ}$  and thus $\Gamma  \hspace{0.25mm}  ( h \to aZ )$ is precisely zero. However, given that the total decay width of the SM Higgs is only about $4 \, {\rm MeV}$, the process $h \to aZ$ can be  important even if deviations from the alignment/decoupling limit are relatively small.

\begingroup 
\renewcommand{\arraystretch}{1.25}
\setlength\tabcolsep{4pt}
\begin{table}[t!]
\centering
\begin{tabular}{c|c|c|c|c}
type & I & II & III & IV \\
\hline 
up-type quarks& $\phantom{-}s_\theta/t_\beta$ & $\phantom{-} s_\theta/t_\beta$ & $\phantom{-} s_\theta/t_\beta$ &$\phantom{-} s_\theta/t_\beta$\\
down-type quarks & $-s_\theta/t_\beta$ & $\phantom{-} s_\theta \hspace{0.25mm} t_\beta$ & $-s_\theta/t_\beta$ & $\phantom{-} s_\theta \hspace{0.25mm} t_\beta$\\
charged leptons & $-s_\theta/t_\beta$ & $\phantom{-} s_\theta \hspace{0.25mm} t_\beta$ & $\phantom{-} s_\theta \hspace{0.25mm} t_\beta$ & $-s_\theta/t_\beta$\\ 
\end{tabular}
\caption{\label{tab:xifM} Ratios $\xi_f^{\rm M}$ of the Yukawa couplings of the pseudoscalar~$a$ relative to those of the SM Higgs in the four types of 2HDM+S models without  tree-level FCNCs.}
\end{table}
\endgroup

Unlike $g_{haZ}$, the triple Higgs coupling $g_{haa}$ depends  not only on the physical Higgs masses and mixing angles but also on some of the trilinear couplings that appear in the full scalar potential. This feature makes the partial decay width $\Gamma  \hspace{0.25mm}  ( h \to aa )$ model dependent, and in consequence the two exotic branching ratios ${\rm BR} \hspace{0.25mm} ( h \to aZ  )$  and ${\rm BR}  \hspace{0.25mm}  ( h \to aa  )$ can  be adjusted freely by  an appropriate choice of parameters. Following this philosophy  we will  treat  ${\rm BR} \hspace{0.25mm}  ( h \to aZ  )$  and ${\rm BR}  \hspace{0.25mm}  ( h \to aa )$ as free parameters in the remainder of this article. 

\section{Numerical results}
\label{sec:results}

We begin our numerical analysis by interpreting the recent CMS results~\cite{Khachatryan:2015wka,Khachatryan:2015nba,Khachatryan:2017mnf} for the exotic SM Higgs decay $h \to aa$ in the 2HDM+S context.  The final states that we consider are $4 \mu$~\cite{Khachatryan:2015wka}, $4 \tau$~\cite{Khachatryan:2015nba,Khachatryan:2017mnf}, $2\mu 2 \tau$~\cite{Khachatryan:2017mnf,CMS-PAS-HIG-17-029}, $2 \mu 2b$~\cite{Khachatryan:2017mnf} and $2 \tau 2 b$~\cite{CMS-PAS-HIG-17-024}. These searches probe $m_a$ values in the range~$[0.25, 3.55] \, {\rm GeV}$, $[4, 8] \, {\rm GeV}$, $[5, 15] \, {\rm GeV}$, $[15, 62.5] \, {\rm GeV}$ and $[25, 62.5] \, {\rm GeV}$, respectively. To facilitate a comparison between the results obtained by  the CMS collaboration and by us, we consider like~\cite{Khachatryan:2017mnf} the following four~2HDM+S benchmark scenarios: the type~I model with $t_\beta = 1$,  the type~II model with $t_\beta = 2$, the  type~III model with $t_\beta = 5$ and the type~IV model with $t_\beta = 0.5$. The  fermionic coupling factors~$\xi_f^{\rm M}$  corresponding to each 2HDM+S type are reported in Table~\ref{tab:xifM}. It is important to realise that the $s_\theta$-dependence of~$\xi_f^{\rm M}$ cancels in ${\rm BR}  \hspace{0.25mm}  ( a \to XX )$ and  it is thus possible to translate constraints on signal strengths such as $\sigma  \hspace{0.25mm}  ( pp \to h )  \hspace{0.5mm}  {\rm BR}  \hspace{0.25mm}  (h \to aa  )  \hspace{0.5mm}  {\rm BR}^2  \hspace{0.25mm}  (a \to \mu^+ \mu^- )$ into $s_\theta$-independent bounds on $\mu_h  \hspace{0.5mm} {\rm BR}  \hspace{0.25mm} (h \to aa )$. Here we have defined $\mu_h = \sigma  \hspace{0.25mm}  ( pp \to h  )/\sigma  \hspace{0.25mm}  ( pp \to h )_{\rm SM}$. 

\begin{figure}[!t]
\begin{center}
\includegraphics[width=0.95\textwidth]{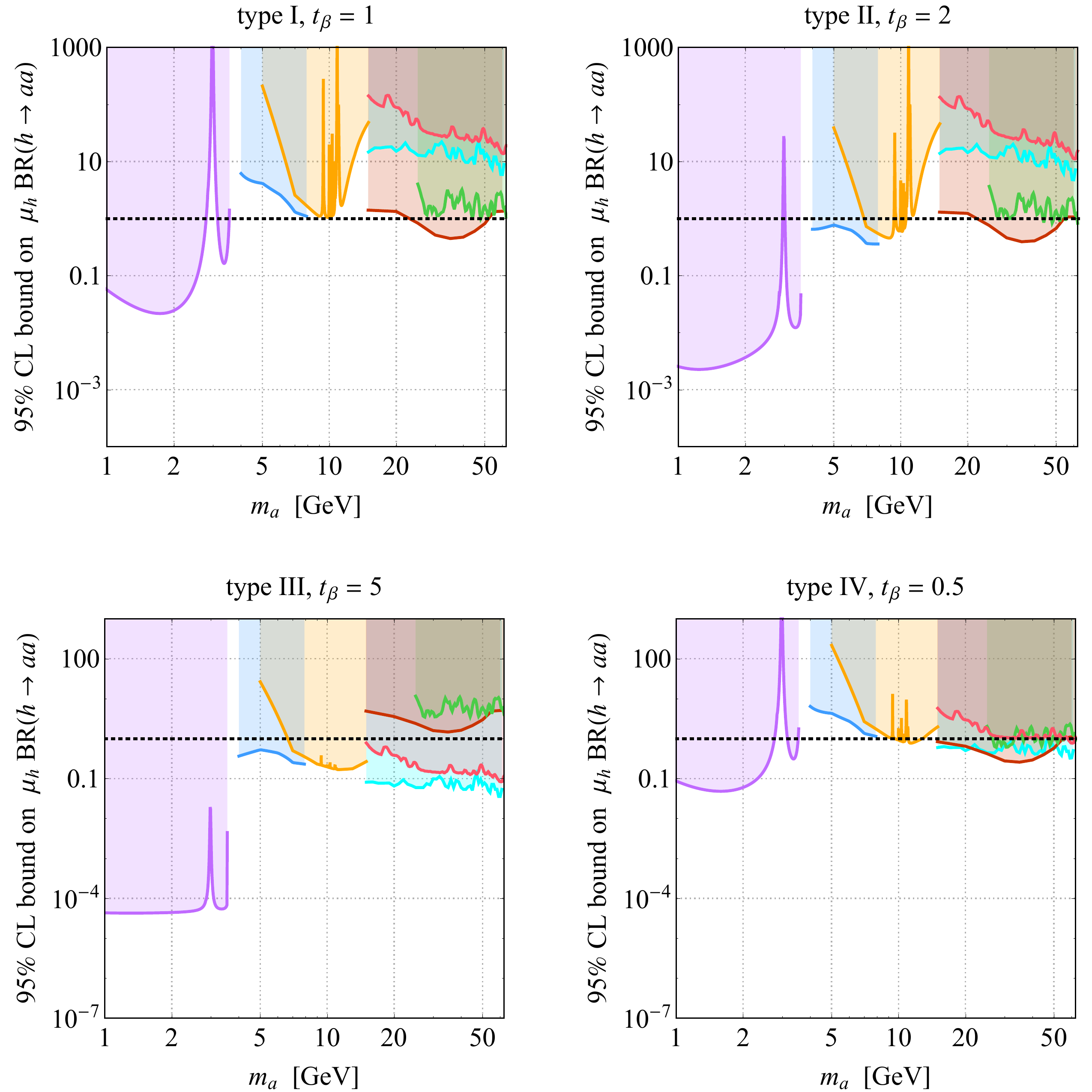}  
\vspace{2mm}
\caption{\label{fig:1} Limits on $\mu_h  \hspace{0.5mm}  {\rm BR}  \hspace{0.25mm} ( h \to aa )$ in the 2HDM+S of type~I with $t_\beta=1$~(top left),  type~II with $t_\beta = 2$~(top right),  type~III with $t_\beta = 5$~(bottom left) and type~IV with $t_\beta = 0.5$ (bottom right). The purple, blue, orange, red, cyan, green and dark red exclusions correspond to the search for $h \to aa \to 4 \mu$~\cite{Khachatryan:2015wka}, $h \to aa \to 4 \tau$~\cite{Khachatryan:2015nba}, $h \to aa \to 4 \tau$~\cite{Khachatryan:2017mnf}, $h \to aa \to 2\mu 2 \tau$~\cite{Khachatryan:2017mnf}, $h \to aa \to 2\mu 2 \tau$~\cite{CMS-PAS-HIG-17-029}, $h \to aa \to 2 \mu 2b$~\cite{Khachatryan:2017mnf} and $h \to aa \to 2 \tau 2 b$~\cite{CMS-PAS-HIG-17-024}, respectively. The dashed black lines indicate $\mu_h  \hspace{0.5mm}  {\rm BR}  \hspace{0.25mm} ( h \to aa ) = 1$ and all coloured regions are excluded at 95\%~CL.
}
\end{center}
\end{figure}

The results of our recast are shown in the  panels of Figure~\ref{fig:1} and should be compared to the  exclusion plots displayed in Figure~8 of~\cite{Khachatryan:2017mnf}. The branching ratios ${\rm BR}  \hspace{0.25mm}  ( a \to XX )$ used to interpret the results in the four particular 2HDM+S scenarios are calculated using the formulas given in Appendix~\ref{app:widths} and include the mixing and threshold effects described in Appendix~\ref{app:mixing}. Notice that the inclusion of~$a \hspace{0.25mm}$--$\hspace{0.5mm} \eta_c$ and $a \hspace{0.25mm}$--$\hspace{0.5mm} \eta_b$ mixing is crucial to obtain meaningful predictions in the $m_a$ regions $[3, 5]~{\rm GeV}$ and $[9, 11] \, {\rm GeV}$, which are left unexplored in the CMS analysis~\cite{Khachatryan:2017mnf}. 

While overall we  observe good agreement between the 95\% confidence level~(CL) exclusions set by CMS and by us, some differences in the derived limits are evident. Firstly, our analysis covers the mass region close to the $c \bar c$ ($b \bar b$) threshold, where our limits display a resonance-like behaviour as a result of the mixing of the $a$ with the three~$\eta_c$ (six $\eta_b$) states included in our study. Second, in the $m_a$ range of~$[1, 3] \, {\rm GeV}$ our bounds on $\mu_h  \hspace{0.5mm}  {\rm BR}  \hspace{0.25mm}  ( h \to aa )$ tend to be somewhat weaker than those derived in~\cite{Khachatryan:2017mnf}.  The observed difference is  again a consequence of the mixing of the $a$ with QCD bound states. In fact, in the very low mass range the total decay width of the unmixed $a$ is below $10^{-3} \, {\rm MeV}$ in the considered~2HDM+S scenarios, while that of the lightest $\eta_c$ state amounts to around $30 \, {\rm MeV}$~\cite{Patrignani:2016xqp}. Hence even a small $\eta_c$-admixture in the mass eigenstate $a$ can lead to an enhanced total decay width $\Gamma_a$ which in turn results in a suppression of ${\rm BR} \hspace{0.25mm} ( a \to \mu^+ \mu^-)$  and a weakening of the bound on~$\mu_h  \hspace{0.5mm} {\rm BR}  \hspace{0.25mm} ( h \to aa )$.  

\begin{figure}[!t]
\begin{center}
\includegraphics[width=0.95\textwidth]{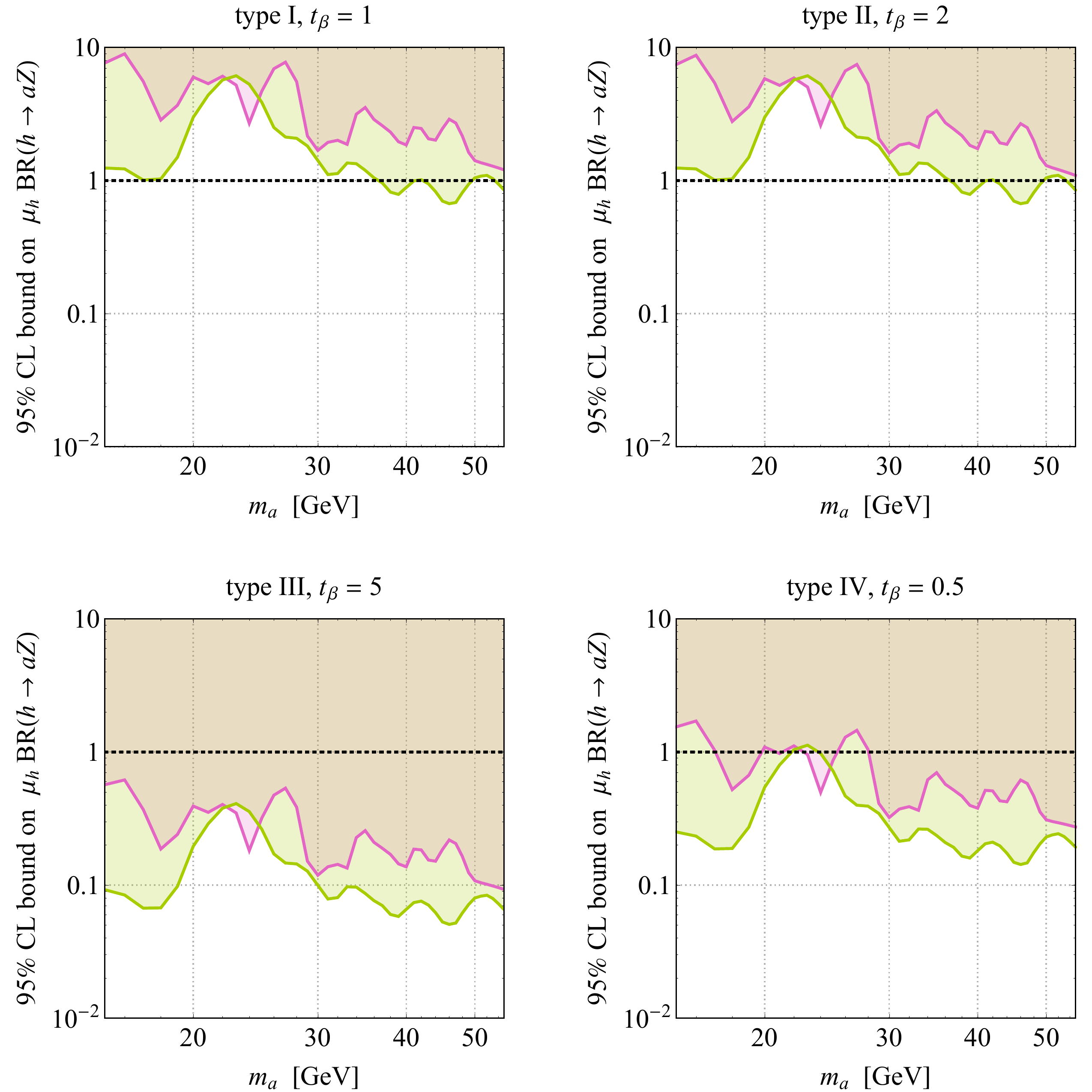}  
\vspace{2mm}
\caption{\label{fig:2} Limits on $\mu_h  \hspace{0.5mm} {\rm BR}  \hspace{0.25mm} ( h \to aZ )$ in the 2HDM+S of type~I with $t_\beta=1$~(top left),  type~II with $t_\beta = 2$~(top right),  type~III with $t_\beta = 5$~(bottom left) and type~IV with $t_\beta = 0.5$ (bottom right). The red and green bounds correspond to the ATLAS search for $pp \to h \to Z_d Z \to 4 \ell$~\cite{Aad:2015sva}  and  $pp \to h \to Z_d Z  \to 2 \mu 2\ell$~\cite{Aaboud:2018fvk}, respectively.   The dashed black lines indicate $\mu_h  \hspace{0.5mm}  {\rm BR}  \hspace{0.25mm} ( h \to aZ ) = 1$ and all coloured regions are excluded at~95\%~CL.}
\end{center}
\end{figure}

A light pseudoscalar $a$ can also be searched for via the decay  $h \to aZ$. The only LHC analyses that presently can be used to set bounds on ${\rm BR} \hspace{0.25mm} ( h \to a Z )$ are the ATLAS searches  for  new dark bosons $Z_d$ produced in $h  \to Z_d Z$~\cite{Aad:2015sva,Aaboud:2018fvk}.  Notice that while the $Z_d$ decays democratically into electrons and muons in the case of the~$a$ one has $\Gamma \hspace{0.25mm} (a \to e^+ e^-)/\Gamma \hspace{0.25mm} (a \to \mu^+ \mu^-) = m_e^2/m_\mu^2 \simeq 2 \cdot 10^{-5}$. As a result $4e$ and $2e2\mu$ events originating from $h \to a Z \to 4e$ and $h \to a Z \to 2e2\mu$ give essentially no contribution to the signal strength in $pp \to h \to aZ \to 4 \ell$. The $8 \, {\rm TeV}$ ATLAS study~\cite{Aad:2015sva} however only provides  exclusion bounds on ${\rm BR} \hspace{0.25mm} (h \to Z_d Z \to 4 \ell)$ from a combination of final states. To correct for this mismatch we have calculated  $r_{{\cal A}\varepsilon} = \sum_{X={4\mu,2\mu2e}} {\cal A}\varepsilon_{X}/\sum_{X={4\mu,4e,2e2\mu,2\mu2e}} {\cal A}\varepsilon_{X}$,  where ${\cal A}\varepsilon_{X}$ denotes the product of acceptance and reconstruction efficiency in the final state $X$ --- the values   for ${\cal A}\varepsilon_{X}$ can be found in the auxiliary material of~\cite{Aad:2015sva}. We find that $r_{{\cal A}\varepsilon}$ has only a mild dependence on $m_a$ and amounts to around $60\%$. The  actual limits are then obtained by equating $r_{{\cal A}\varepsilon}   \,  {\rm BR} \hspace{0.25mm}  (h \to aZ)   \,  {\rm BR} \hspace{0.25mm} (a \to \mu^+ \mu^-) \, {\rm BR} \hspace{0.25mm} (Z \to \ell^+ \ell^-)  = {\rm BR} \hspace{0.25mm} (h \to Z_d Z \to 4 \ell)$ and solving for ${\rm BR} \hspace{0.25mm}  (h \to aZ)$. To~improve upon this naive recast one would need   individual bounds for the different  combinations of final-state lepton flavours. In fact,  the very recent $13 \, {\rm TeV}$ ATLAS analysis~\cite{Aaboud:2018fvk} provides ${\cal A}\varepsilon_{2\mu 2 \ell}$ as well as limits on the relevant fiducial  cross section. Our recast of the latter results thus only has to rely on the assumption that  the product ${\cal A}\varepsilon_{2\mu 2 \ell}$ is roughly the same for the  $Z_d$ model and the 2HDM+S scenario, which we indeed believe to be the case. 

\newpage 

The exclusion limits  on $\mu_h  \hspace{0.5mm} {\rm BR}  \hspace{0.25mm} ( h \to aZ )$ corresponding to the four  2HDM+S benchmark scenarios discussed earlier are presented in Figure~\ref{fig:2}. From the panels it is evident that, apart from pseudoscalar masses around $25 \, {\rm GeV}$ where the data~\cite{Aaboud:2018fvk} has a local deficit, the constraints that derive from the $13 \, {\rm TeV}$ analysis~\cite{Aaboud:2018fvk}  are significantly stronger than those that one obtains from the $8 \, {\rm TeV}$ data~\cite{Aad:2015sva}. One also observes that the constraints in the first and second benchmark are weak as they just start to probe the region $\mu_h  \hspace{0.5mm} {\rm BR} \hspace{0.25mm} (h \to aZ) \lesssim 1$, whereas in the third and fourth 2HDM+S scenario already values of $\mu_h  \hspace{0.5mm} {\rm BR} \hspace{0.25mm} (h \to aZ) \lesssim 0.1$ can be probed with the available LHC data sets. Since the  asymmetry between electron and muon  final states from $h \to aZ$ decays is a striking signature of a light pseudoscalar, we strongly encourage  our experimental colleagues to provide as in~\cite{Aaboud:2018fvk} separate bounds for the  $2e2\ell$ and $2\mu2\ell$ final states in future searches for   signatures of the type  $h \to Z_d Z \to 4 \ell$. 

\begin{figure}[!t]
\begin{center}
\includegraphics[width=0.95\textwidth]{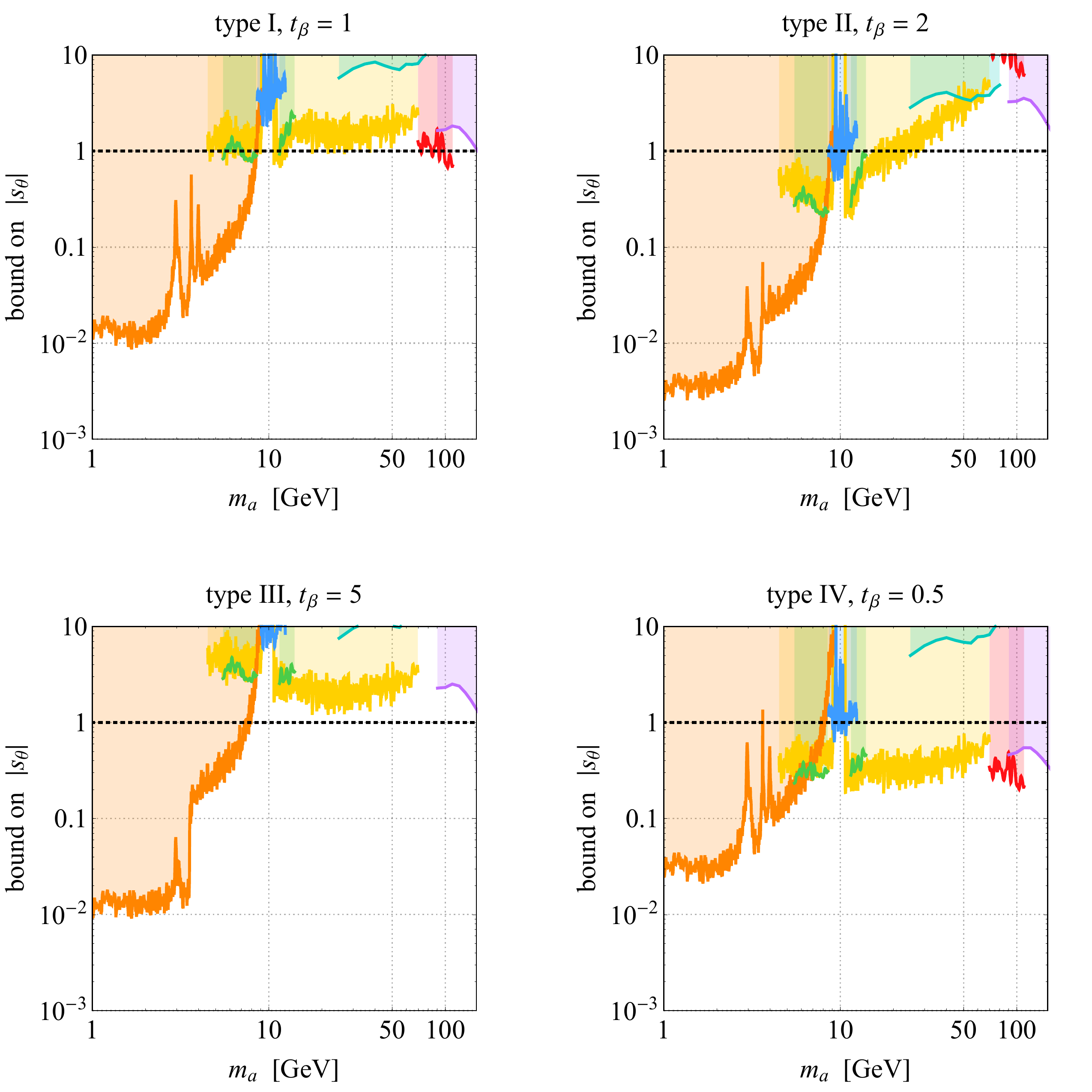}
\vspace{2mm}
\caption{\label{fig:3}   Limits on $|s_\theta|$ in the 2HDM+S of type~I with $t_\beta = 1$~(top left),  type~II with $t_\beta = 2$~(top right),  type~III with $t_\beta = 5$~(bottom left) and type~IV with $t_\beta = 0.5$ (bottom right). The green, turquoise, red, purple, orange, blue and yellow exclusions correspond to the searches for $a \to \mu^+ \mu^-$~\cite{Chatrchyan:2012am}, $pp \to a b \bar b \to\tau^+ \tau^-  b \bar b$~\cite{Khachatryan:2015baw}, $pp \to a \to \gamma \gamma$~\cite{CMS-PAS-HIG-17-013}, $pp \to a \to \tau^+ \tau^-$~\cite{CMS-PAS-HIG-16-037} and $\Upsilon (1S) \to a \gamma \to \mu^+ \mu^- \gamma$~\cite{Lees:2012iw},  the measurements of $\Upsilon$~production~\cite{Haisch:2016hzu,Aaij:2015awa} and the inclusive dimuon cross section~\cite{Aaij:2017rft}, respectively. The dashed black lines indicate $|s_\theta| = 1$ and all coloured regions are excluded at 95\%~CL apart from the orange and yellow contours which only hold at 90\%~CL. 
}
\end{center}
\end{figure}

Constraints on the parameter space of the four different types of 2HDM+S scenarios can finally  be derived from the LHC searches for $pp \to a \to \mu^+ \mu^-$~\cite{ATLAS:2011cea,Chatrchyan:2012am}, $pp \to a b \bar b \to\tau^+ \tau^-  b \bar b$~\cite{Khachatryan:2015baw} or $pp \to a b \bar b \to\mu^+ \mu^-  b \bar b$~\cite{Sirunyan:2017uvf}, $pp \to a \to \gamma \gamma$~\cite{CMS-PAS-HIG-17-013}, $pp\to a \to \tau^+ \tau^-$~\cite{CMS-PAS-HIG-16-037}, from the studies of  $\Upsilon  \to a \gamma$ decays performed at BaBar~\cite{Lees:2011wb,Lees:2012iw,Lees:2012te} and from the LHCb measurements of $\Upsilon$  production~\cite{Haisch:2016hzu,Aaij:2015awa} as well as of the inclusive dimuon cross section~\cite{Ilten:2016tkc,Aaij:2017rft}. Since these search strategies all rely on the production of a pseudoscalar $a$ the resulting constraints all scale as $s_\theta^2$. For a given type of 2HDM+S model and a fixed value of $t_\beta$, the measurements~\cite{ATLAS:2011cea,Chatrchyan:2012am,Khachatryan:2015baw,Sirunyan:2017uvf,CMS-PAS-HIG-17-013,CMS-PAS-HIG-16-037,Lees:2011wb,Lees:2012iw,Lees:2012te,Aaij:2017rft} can therefore be used to set limits on $|s_\theta|$ as a function of~the pseudoscalar mass $m_a$. 

For concreteness we study the same four 2HDM+S  scenarios that we have already considered before. The  most stringent limits on $|s_\theta|$ that can be derived at present  are displayed in Figure~\ref{fig:3}. In order to~recast the results of the CMS searches for $a \to \mu^+ \mu^-$~\cite{Chatrchyan:2012am},  $pp \to a b \bar b \to\tau^+ \tau^-  b \bar b$~\cite{Khachatryan:2015baw}, $pp \to a \to \gamma \gamma$~\cite{CMS-PAS-HIG-17-013}, $pp \to a \to \tau^+ \tau^-$~\cite{CMS-PAS-HIG-16-037}, the LHCb measurements of $\Upsilon$  production~\cite{Haisch:2016hzu,Aaij:2015awa} and the inclusive dimuon cross section~\cite{Aaij:2017rft}, one needs to know the production cross sections of a light $a$ in gluon-fusion and in association with $b \bar b$ pairs.  Our predictions for~$gg \to a$ production are obtained  at next-to-next-to-leading order in QCD using {\tt HIGLU}~\cite{Spira:1995mt}, while the $pp \to a b \bar b$ cross sections are calculated at next-to-leading order~(NLO) in QCD in the four-flavour scheme with {\tt MadGraph5\_aMCNLO}~\cite{Alwall:2014hca} employing an {\tt UFO} implementation~\cite{Degrande:2011ua} of the 2HDM model discussed in the publication~\cite{Bauer:2017ota}. 

Our recast of the results of the LHCb search for dark photons $A^\prime$~\cite{Aaij:2017rft} proceeds as follows. We calculate the inclusive $pp \to A^\prime$ production cross section at NLO in QCD with the help of {\tt MadGraph5\_aMCNLO}~\cite{Alwall:2014hca}, while we extract ${\rm BR} \hspace{0.25mm} (A^\prime \to \mu^+ \mu^-)$ from the well-measured cross section ratio~$R = \sigma \hspace{0.25mm} (e^+ e^- \to {\rm hadrons})/\sigma \hspace{0.25mm} (e^+ e^- \to \mu^+ \mu^-)$~\cite{Patrignani:2016xqp}. Following~\cite{Ilten:2016tkc,Aaij:2017rft}, model-dependent $A^\prime \hspace{0.25mm}$--$\hspace{0.5mm} Z$ mixing effects are included in our calculation employing the formulas given in~\cite{Cline:2014dwa}. We have also  taken into account detector acceptance differences between  $pp \to A^\prime \to \mu^+ \mu^-$  and $pp \to a  \to \mu^+ \mu^-$ by computing the ratio $r_{\cal A} = {\cal A}_a/{\cal A}_{A^\prime}$ of signal acceptances.   We find that $r_{\cal A}$ amounts to around 2.0, 1.3, 1.0 at $m_a = 5 \, {\rm GeV}, 15 \, {\rm GeV}, 70 \, {\rm GeV}$ and scales approximately linear between the quoted $m_a$ values. Concerning the detection efficiencies ${\varepsilon}_{A^\prime}$ and ${\varepsilon}_{a}$ we assume that they are identical for $A^\prime \to \mu^+ \mu^-$ and $a \to \mu^+ \mu^-$, which should be a good approximation when the dimuon signal is prompt~\cite{Aaij:2017rft}. We finally add that in our recast of the LHCb dark photon results, we only consider the mass region $m_a > 4.5 \, {\rm GeV}$  to avoid $a \hspace{0.25mm}$--$\hspace{0.5mm} \eta_c$ mixing contributions to the $pp \to a$ cross section associated to $pp \to \eta_c$ production. The mass region $m_a \in [9.1, 10.6] \, {\rm GeV}$ is also not covered by our recast, because in~\cite{Aaij:2017rft} the LHCb collaboration does not present bounds on the kinetic mixing of the $A^\prime$ close to the $b \bar b$  threshold.

The main conclusion  that can be drawn from the results presented in  Figure~\ref{fig:3} is that only in the 2HDM+S scenario of  type IV with $t_\beta = 0.5$ it is possible to set physical meaningful bounds on the sine of the mixing angle $\theta$,~i.e.~$|s_\theta| < 1$, over the entire range of studied pseudoscalar masses. One furthermore observes that solely the BaBar search for the radiative decay $\Upsilon (1S) \to a \gamma \to \mu^+ \mu^- \gamma$~\cite{Lees:2012iw}  allows to probe parameter regions with $|s_\theta| < 0.1$. This search is however kinematically limited to $m_a < m_{\Upsilon(1S)} \simeq 9.5 \, {\rm GeV}$.  Improvements in the existing LHC search strategies (and/or new approaches) are needed to reach the same sensitivity on~$|s_\theta|$ for pseudoscalar masses above approximately $10 \, {\rm GeV}$ in the examined 2HDM+S benchmark models. Measurements of the inclusive dimuon cross section~\cite{Ilten:2016tkc,Aaij:2017rft} seem to be quite promising in this context. 

\section{Conclusions}
\label{sec:conclusions}

Beyond the SM theories with an extended Higgs sector can naturally lead to pseudoscalar resonances with masses significantly below the EW scale if these states serve as PNGBs of  an approximate global  $U(1)$ symmetry. The $R$-symmetry limit in the NMSSM and the case of spontaneously broken $U(1)$ subgroups in Little Higgs models are just two working examples of this general idea. Searches for light CP-odd spin-0 states are thus theoretically well-motivated and in the case of a detection could help to illuminate the structure and dynamics of the underlying UV model. 

The existing  collider searches for pseudoscalars with masses  of approximately $[1, 100]~{\rm GeV}$ fall into two different classes.  Firstly, searches that look for the presence of a light~$a$ in the decay of a SM particle. Searches for $h \to aa$ and $h \to aZ$, but also the radiative decays $\Upsilon \to a \gamma$ belong to this category.  In the case of the exotic Higgs decays the resulting signature that the ATLAS and CMS experiments have explored  are four-fermion final states containing at least two opposite-sign leptons~\cite{Khachatryan:2015wka,CMS:2016tgd,Khachatryan:2015nba,Khachatryan:2017mnf,CMS-PAS-HIG-17-029,CMS-PAS-HIG-17-024,Aad:2015sva,Aaboud:2018fvk}, while what concerns the radiative~$\Upsilon$~decays, BaBar has considered the hadronic, dimuon and ditau decays of pseudoscalars~\cite{Lees:2011wb,Lees:2012iw,Lees:2012te}.  The second type of searches instead relies on the direct production of the $a$ in $pp$ collisions and its subsequent decays to either charged lepton or photon pairs. Both the gluon-fusion channel~\cite{Haisch:2016hzu,ATLAS:2011cea,Chatrchyan:2012am,CMS-PAS-HIG-17-013,Mariotti:2017vtv,CMS-PAS-HIG-16-037,Aaij:2015awa} and $ab \bar b$ production~\cite{Khachatryan:2015baw,Sirunyan:2017uvf} have so far been exploited to look for light pseudoscalars at the LHC in this way. 

In this work, we have performed a global analysis of the present collider constraints on light pseudoscalar states. To facilitate  a comparison with the recent CMS study~\cite{Khachatryan:2017mnf}, we have considered the class of 2HDM+S models, treating the parameters $t_\beta$ and $s_\theta$ as well as the branching ratios ${\rm BR} \hspace{0.25mm} ( h \to aa )$ and ${\rm BR} \hspace{0.25mm} ( h \to aZ )$ as free parameters --- see Section~\ref{sec:generalities} for a concise introduction to the  2HDM+S setup. A complication that arises in our analysis is that in the mass regions $[3, 5] \, {\rm GeV}$ and $[9,11] \, {\rm GeV}$, non-perturbative effects such as the mixing of the pseudoscalar with QCD bound states have to be taken into account to allow for a meaningful interpretation of the experimental data. We have worked out the  theoretical formalism necessary to calculate the  most relevant short-distance and long-distance effects and provide a collection of the corresponding formulas in the two Appendices~\ref{app:widths} and~\ref{app:mixing}.  

Our numerical analysis consists of three parts. In the first part, we have derive 95\%~CL exclusion limits  on the signal strength $\mu_h  \hspace{0.5mm} {\rm BR}  \hspace{0.25mm} (h \to aa )$ that follow from the latest CMS searches for the  exotic $h \to aa$ decay~\cite{Khachatryan:2015wka,CMS:2016tgd,Khachatryan:2015nba,Khachatryan:2017mnf,CMS-PAS-HIG-17-029,CMS-PAS-HIG-17-024}, while in the second part we present the limits on $\mu_h  \hspace{0.5mm} {\rm BR}  \hspace{0.25mm} (h \to aZ )$ that stem from the ATLAS searches for $h \to Z_d Z \to 4 \ell$~\cite{Aad:2015sva} and $h \to Z_d Z \to 2\mu 2 \ell$~\cite{Aaboud:2018fvk}. The exclusion bounds on~$|s_\theta|$ that arise from the searches~\cite{Haisch:2016hzu,Chatrchyan:2012am,Khachatryan:2015baw,CMS-PAS-HIG-17-013,CMS-PAS-HIG-16-037,Lees:2012iw,Aaij:2015awa,Aaij:2017rft} are finally derived in the third part of our numerical study. In all three cases, we have considered four specific 2HDM+S benchmark scenarios that differ in the choice of Yukawa sector and $t_\beta$.  We have found that the inclusion of $a \hspace{0.25mm}$--$\hspace{0.5mm} \eta_c$ $\big($$a \hspace{0.25mm}$--$\hspace{0.5mm} \eta_b$$\big)$ mixing effects as well as open flavour decays to~$D$~$\big ($$B_{(s)}$$\big )$ meson pairs has a visible impact on the obtained limits only in the mass region of approximately $[1, 4] \, {\rm GeV}$ $\big ($$[10, 15] \, {\rm GeV}$$\big )$, while   perturbative calculations are perfectly adequate  for~$m_a$~values away from the~$c\bar c$ and $b \bar b$ thresholds. 

The main conclusion that can be drawn from the results presented in Figures~\ref{fig:1},~\ref{fig:2} and~\ref{fig:3} is that existing collider constraints on the parameter space of 2HDM+S models are in general not very strong. Exceptions are the  $[1, 3] \, {\rm GeV}$ region in which $\mu_h \hspace{0.25mm} {\rm BR} \hspace{0.25mm} ( h \to a a )$ is well-constrained by the CMS search for $h \to aa \to 4 \mu$~\cite{Khachatryan:2015wka} and the $[1, 9.5] \, {\rm GeV}$ range where the $\Upsilon(1S) \to a \gamma \to \mu^+ \mu^- \gamma$ search of BaBar~\cite{Lees:2012iw} provides stringent limits on $|s_\theta|$. Much to the opposite, the 2HDM+S parameter space turns out to be least constrained for $m_a $ values in the range of approximately $ [15, 70] \, {\rm GeV}$. The development of improved or new search techniques (such as for instance dedicated searches for $h \to a Z$~\cite{Aaboud:2018fvk}~and inclusive diphoton~\cite{Mariotti:2017vtv} or dimuon~\cite{Ilten:2016tkc,Aaij:2017rft} cross section measurements) that specifically focus  on the latter mass region therefore seems to be a worthwhile scientific goal. 

\acknowledgments 
We are grateful to Kai~Schmidt-Hoberg for providing details on the estimate of $\Gamma  \hspace{0.25mm}  ( a \to KK\pi  )$ as given in~\cite{Dolan:2014ska}. We furthermore thank Ulrich~Ellwanger, Filippo~Sala, Dominik~St{\"o}ckinger and Mike~Williams for their interest in our work, constructive feedback and their useful suggestions. UH appreciates the continued hospitality  and support of the CERN Theoretical Physics Department. JFK acknowledges the financial support from the Slovenian Research Agency (research core funding No.~P1-0035 and J1-8137). MS~would like to thank the organisers of Les~Houches for the great and fruitful atmosphere of the workshop.

\begin{appendix}

\section{Decay width formulas}
\label{app:widths}

In the calculation of the total decay  width $\Gamma_a$ of the unmixed pseudoscalar $a$, we employ the following expressions for the partial decay widths~(see the reviews~\cite{Spira:1997dg,Djouadi:2005gi,Djouadi:2005gj,Spira:2016ztx} for instance)
\begin{align} 
\Gamma\hspace{0.25mm} (a\to \ell^+ \ell^-) & = \frac{\big ( \xi_\ell^{\rm M}  \big )^2  \hspace{0.35mm}  m_\ell^2 \hspace{0.25mm}  m_a}{8\pi v^2} \hspace{0.5mm}   \beta_{\ell/a}  \,, \label{eq:A1} \\[1mm]
\Gamma\hspace{0.25mm} (a\to q\bar q) & = \frac{3 \hspace{0.25mm} \big (\xi_q^{\rm M} \big )^2  \hspace{0.25mm} \overline m_q^2 \hspace{0.25mm}  m_a}{8\pi v^2} \left  (1+\Delta_q + \frac{\xi_t^{\rm M} }{\xi_q^{\rm M}} \hspace{0.25mm} \Delta_t \right )\,, \label{eq:A2} \\[1mm]
\Gamma\hspace{0.25mm} (a\to Q\bar Q) & = \frac{3 \hspace{0.25mm} \big (\xi_Q^{\rm M} \big )^2   \hspace{0.25mm} m_Q^2 \hspace{0.25mm}  m_a}{8\pi v^2} \hspace{0.5mm}  \beta_{Q/a} \left  (1+\Delta_Q  \right )\,, \label{eq:A3} \\[1mm]
\Gamma\hspace{0.25mm} (a\to gg) & = \frac{\alpha_s^2 \hspace{0.25mm} m_a^3}{32\pi^3 v^2} \left| \sum_{q=t,b,c,s} \xi_q^{\rm M}  \hspace{0.5mm}  
\mathcal P (\tau_{q/a}) \right|^2  \hspace{0.25mm} K_g  \,, \label{eq:A4} \\[1mm]
\Gamma\hspace{0.25mm} (a\to \gamma\gamma) & = \frac{\alpha^2 \hspace{0.25mm} m_a^3}{64\pi^3 v^2} \left| \sum_{q=t,b,c,s} 3 \hspace{0.25mm} \xi_q^{\rm M}  Q_q^2 \hspace{0.5mm}  \big ( \mathcal P ( \tau_{q/a})+  \Delta_\gamma \big ) +  \xi_\tau^{\rm M}   \hspace{0.25mm} \mathcal P ( \tau_{\tau/a}) \right|^2 \,, \label{eq:A5} 
\end{align}
where $\overline{\rm MS}$ masses are indicated by a bar while masses without a bar are evaluated in the pole scheme.  We  have furthermore defined $\tau_{f/a} = 4  m_{f} ^2/m_a^2$ and $\beta_{f/a} = \sqrt{1 - \tau_{f/a}}$ and used the symbol~$Q_q$ to denote the electric charge of the quark in question. All $\overline{\rm MS}$ masses  as well as  the coupling constants $\alpha_s$ and $\alpha$ are renormalised at the scale $\mu_R = m_a$.  Table~\ref{tab:xifM} finally contains the coupling assignments~$\xi_f^{\rm M}$ that we consider in our work.  

The  QCD corrections to the partial decay width into light quarks~(\ref{eq:A2}) that are included in our numerical analysis read~\cite{Drees:1989du,Braaten:1980yq,Sakai:1980fa,Inami:1980qp,Gorishnii:1983cu,Drees:1990dq,Gorishnii:1990zu,Gorishnii:1991zr,Kataev:1993be,Surguladze:1994gc,Melnikov:1995yp,Chetyrkin:1996sr}
\begin{equation} \label{eq:A7}
\begin{split}
\Delta_q & = \frac{\alpha_s}{\pi}  \hspace{0.75mm}  5.67  + \left ( \frac{\alpha_s}{\pi}  \right  )^2 \hspace{0.25mm} \big ( 35.94 - 1.36 \hspace{0.25mm}  N_f \big )  +  \left ( \frac{\alpha_s}{\pi}  \right )^3  \Big (164.14 - 25.77 \hspace{0.25mm} N_f + 0.259  \hspace{0.25mm} N_f^2 \Big )   \\[1mm]
& \phantom{xx} + \left ( \frac{\alpha_s}{\pi}  \right )^4 \hspace{0.25mm} \Big ( 39.34 - 220.9  \hspace{0.25mm} N_f  + 9.685 \hspace{0.25mm} N_f^2  - 0.0205 \hspace{0.25mm} N_f^3 \Big )  \,,
\end{split}
\end{equation} 
and~\cite{Chetyrkin:1995pd,Larin:1995sq}
\begin{equation}  \label{eq:A8}
\Delta_t  =  \left ( \frac{\alpha_s}{\pi}   \right )^2 \, \left [ 3.83 + \ln \left ( \frac{m_t^2}{m_a^2} \right ) + \frac{1}{6} \ln^2  \left ( \frac{\overline m_q^2}{m_a^2} \right )  \right ] \,.
\end{equation}
The symbol $N_f$  introduced above denotes the number of light quark flavours that are active at the scale~$m_a$.  For pseudoscalar masses far above the threshold, i.e. $m_a \gg 2 m_q$, the results~(\ref{eq:A7}) and~(\ref{eq:A8}) represent at the moment the most accurate predictions for the QCD corrections to $\Gamma \hspace{0.25mm} ( a \to q \bar q)$. In our numerical analysis, we hence use them to calculate the partonic rate of $a \to s \bar s$.  

In  the case of the partial decay width  into heavy-quark pairs~(\ref{eq:A3}) the  QCD corrections  are given to first order in $\alpha_s$ by ~\cite{Drees:1989du,Braaten:1980yq,Sakai:1980fa,Inami:1980qp,Gorishnii:1983cu,Drees:1990dq} 
\begin{equation} \label{eq:A9}
\Delta_Q  =  \frac{\alpha_s}{\pi} \left ( \frac{4 {\cal Q} (\beta_{Q/a})}{3 \beta_{Q/a}}  - \frac{19+2\beta_{Q/a}^2+3 \beta_{Q/a}^4 }{12 \beta_{Q/a}} \ln x_{\beta_{Q/a}} + \frac{21  - 3 \beta_{Q/a}^2}{6} \right ) \,.
\end{equation}
Here we have introduced the abbreviation $x_{\beta_{Q/a}} = (1 - \beta_{Q/a})/(1 + \beta_{Q/a})$ and  the one-loop function entering~(\ref{eq:A9}) takes the form 
\begin{equation}
{\cal Q} (\beta) = \big (1+\beta^2 \big )\left ( 4 \hspace{0.25mm} {\rm Li}_2 (x_\beta) + 2  \hspace{0.25mm}  {\rm Li}_2 (-x_\beta) + 4 \ln x_\beta \ln \frac{2}{1+\beta} + 2 \ln x_\beta \ln \beta \right ) - 3 \beta \ln \frac{4 \beta^{4/3}}{1-\beta^2} \,,  \label{eq:A10}
\end{equation}
with ${\rm Li}_2 (z)$ denoting the usual dilogarithm. In the threshold region,~i.e.~$m_a \simeq 2 m_Q$, mass effects are important and as a result  the QCD corrections~(\ref{eq:A9}) should be used to describe them.  Following the prescription implemented in {\tt HDECAY}~\cite{Djouadi:1997yw,Djouadi:2018xqq}, the transition between the region close to threshold to that far above threshold is  achieved by a smooth linear interpolation of the results~(\ref{eq:A2}) and~(\ref{eq:A3}).  Because this approach yields an optimised description of $\Gamma \hspace{0.25mm} ( a \to c \bar c )$ $\big(\Gamma \hspace{0.25mm} ( a \to b \bar b )\big)$ for pseudoscalar masses in the vicinity of $m_a \simeq 3.1 \, {\rm GeV}$ ($m_a \simeq 11.5 \, {\rm GeV}$) it is used in our work.  

The one-loop function appearing in~(\ref{eq:A4}) and~(\ref{eq:A5}) is given by 
\begin{equation} \label{eq:A11}
\mathcal P (\tau)  = \tau \arctan^2 \left( \frac{1}{\sqrt{\tau-1}} \right) \,,
\end{equation}
where for analytic continuation it is understood that  $\tau \to \tau - i 0$. 

The multiplicative factor $K_g$ entering (\ref{eq:A4})  takes the following form 
\begin{equation} \label{eq:Kg}
K_g = 1 + 2 \hspace{0.25mm} {\rm Re} \left ( \frac{\sum_{q=t,b,c,s}  \hspace{0.25mm} \xi_q^{\rm M} \hspace{0.5mm}  \Delta_g}{ \sum_{q=t,b,c,s}  \hspace{0.25mm}  \xi_q^{\rm M} \hspace{0.5mm} {\cal P}  (\tau_{q/a}) } \right ) + \frac{\alpha_s}{\pi} \left ( \frac{73}{4} - \frac{7}{6} \hspace{0.25mm} N_f  \right ) \,,
\end{equation}
where the second term encodes the virtual two-loop QCD corrections, while the third term corresponds to the finite part of the real QCD corrections in the heavy-quark limit~\cite{Spira:1997dg,Spira:1995rr}. We have verified that quark mass effects of the real corrections not included in (\ref{eq:Kg}) amount to no more than $5\%$.  The virtual corrections can be written as 
\begin{equation} \label{eq:A12}
\Delta_g  =  \frac{\alpha_s}{\pi}  \left(  {\cal G} (y_{q/a})+  2 \hspace{0.25mm} \tau_{q/a} \hspace{0.25mm}  {\cal P}^\prime (\tau_{q/a}) \, \ln \frac{\mu_q^2}{m_q^2}  \right)\,,
\end{equation}
where $y_{q/a}= -x_{q/a}$ with $\tau_{q/a} \to \tau_{q/a} + i 0$ for analytic continuation and the prime denotes a derivative with respect to $\tau_{q/a}$. To reproduce the position of the $a \to q \bar q$ threshold correctly, we set~$\mu_q = m_a/2$ in our study.  The two-loop function appearing in~(\ref{eq:A12}) reads~\cite{Spira:1995rr,Harlander:2005rq}
\begin{align}
{\cal G} (y) & = \frac{y}{ {\left(1-y\right)}^2}\biggl[ 48 \hspace{0.25mm} {\rm H}(1,0,-1,0;y)+ 4\ln(1-y)\ln^3 y- 24\hspace{0.25mm} \zeta_2 \hspace{0.25mm} {\rm Li}_2( y)- 24 \hspace{0.25mm} \zeta_2 \ln(1- y)\ln y  \nonumber \\[1mm]
& \hspace{1.8cm} - 72 \hspace{0.25mm}\zeta_3 \ln(1- y) - \frac{220}{3} \hspace{0.25mm} {\rm Li}_3( y) - \frac{128}{3} \hspace{0.25mm} {\rm Li}_3(- y) + 68 \hspace{0.25mm} {\rm Li}_2( y)\ln y \nonumber \\[1mm] 
& \hspace{1.8cm} + \frac{64}{3} \hspace{0.25mm} {\rm Li}_2(- y)\ln y + \frac{94}{3}\ln(1- y)\ln^2 y  - \frac{16}{3}\hspace{0.25mm}\zeta_2\ln y + \frac{124}{3}\hspace{0.25mm}\zeta_3 + 3\ln^2 y \biggr] \nonumber \\[1mm]
& \phantom{xx} - \frac{ 24  y \left(5+7 { y}^2\right) }{{\left(1- y\right)}^3 \left(1+ y\right)} \hspace{0.25mm} {\rm Li}_4( y)  -\frac{ 24 y \left(5+11 { y}^2\right)}{{\left(1- y\right)}^3 \left(1+ y\right)} \hspace{0.25mm} {\rm Li}_4(-y)   \nonumber  \\[1mm]
&  \phantom{xx} +\frac{ 8 y \left(23+41 { y}^2\right) } {3{\left(1- y\right)}^3 \left(1+ y\right)} \biggl[ {\rm Li}_3( y) +{\rm Li}_3(- y) \biggr] \ln y -\frac{ 4 y \left(5+23 { y}^2\right) }{3{\left(1- y\right)}^3 \left(1+ y\right)}\hspace{0.25mm} {\rm Li}_2( y)\ln^2 y  \nonumber \\[1mm] 
&  \phantom{xx} - \frac{ 32 y \left(1+{ y}^2\right) }{3{\left(1- y\right)}^3 \left(1+ y\right)} \hspace{0.25mm} {\rm Li}_2(- y)\ln^2 y  +\frac{  y \left(5-13 { y}^2\right) }{36 {\left(1- y\right)}^3 \left(1+ y\right)}\ln^4 y +\frac{ 2 y \left(1-17 { y}^2\right) }{3{\left(1- y\right)}^3 \left(1+ y\right)}\hspace{0,25mm} \zeta_2\ln^2 y \hspace{4mm} \nonumber \\[1mm] 
&  \phantom{xx} +\frac{ 4 y \left(11-43 { y}^2\right) }{3{\left(1- y\right)}^3 \left(1+ y\right)} \hspace{0.25mm} \zeta_3\ln y +\frac{ 24 y \left(1-3 { y}^2\right) }{{\left(1- y\right)}^3 \left(1+ y\right)} \hspace{0.25mm} \zeta_4 +\frac{ 2 y \left(2+11 y\right) }{ 3{\left(1- y\right)}^3}\ln^3 y \,.
\end{align}
Here ${\rm{H}}(1,0,-1,0;z)$ is a  harmonic polylogarithm of weight four  with two indices different from zero, which we evaluate numerically with the help of the program {\tt HPL}~\cite{Maitre:2007kp}. The polylogarithm of order three~(four) is denoted by ${\rm Li}_3 (z)$~$\big($${\rm Li}_4 (z)$$\big)$, while  $\zeta_2 = \pi^2/6$, $\zeta_3 \simeq 1.20206$ and $\zeta_4 = \pi^4/90$ are the relevant  Riemann's zeta values. 

In the case of (\ref{eq:A5}) we decompose the relevant QCD corrections as 
\begin{align} \label{eq:Deltagamma}
\Delta_\gamma & =  \frac{\alpha_s}{\pi}  \left(  {\cal A} (y_{q/a})+  2 \hspace{0.25mm} \tau_{q/a} \hspace{0.25mm}  {\cal P}^\prime (\tau_{q/a}) \, \ln \frac{\mu_q^2}{m_q^2}  \right)\,,
\end{align}
with~\cite{Spira:1995rr,Harlander:2005rq,Aglietti:2006tp}
\begin{align}
{\cal A} (y) & = - \frac{ y \left(1+y^2\right) }{{\left(1-y\right)}^3 (1+y)}  \biggl[ 72 \hspace{0.25mm} {\rm Li}_4(y) + 96 \hspace{0.25mm} {\rm Li}_4(-y) - \frac{128}{3} \hspace{0.25mm} \big[  {\rm Li}_3(y) + {\rm Li}_3(-y) \big ] \ln y  \nonumber \\[1mm] 
& \hspace{3.05cm} + \frac{28}{3} \hspace{0.25mm}  {\rm Li}_2(y)\ln^2 y + \frac{16}{3} \hspace{0.25mm}  {\rm Li}_2(-y)\ln^2 y + \frac{1}{18}\ln^4 y \nonumber \\[1mm]
& \hspace{3.05cm} + \frac{8}{3} \hspace{0.25mm}\zeta_2\ln^2 y + \frac{32}{3}\hspace{0.25mm} \zeta_3\ln y + 12\hspace{0.25mm}\zeta_4 \biggr]  \\[1mm] 
& \phantom{xx}  +\frac{y }{{\left(1-y\right)}^2} \biggl[ -\frac{56}{3} \hspace{0.25mm}  {\rm Li}_3(y) - \frac{64}{3}\hspace{0.25mm}   {\rm Li}_3(-y) + 16 \hspace{0.25mm}  {\rm Li}_2 (y) \ln y + \frac{32}{3} \hspace{0.25mm}  {\rm Li}_2(-y)\ln y  \nonumber \\[1mm] 
& \hspace{2.15cm} +\frac{20}{3} \ln \left ( 1 - y \right ) \ln^2 y -\frac{8}{3} \hspace{0.25mm}\zeta_2\ln y + \frac{8}{3} \hspace{0.25mm} \zeta_3 \biggr] +\frac{2y \left(1+y\right) } {3 {\left(1-y\right)}^3} \ln^3 y \,.  \nonumber 
\end{align}

\section{Mixing and threshold effects}
\label{app:mixing}

Even though the decay $a \to b \bar b$ ($a \to c \bar c$) is kinematically forbidden below the open-flavour threshold, the presence of heavy quarks can become relevant through mixing between the pseudoscalar~$a$ and bottomonium (charmonium) bound states with the same quantum numbers~\cite{Drees:1989du,Domingo:2008rr,Domingo:2010am,Domingo:2011rn,Baumgart:2012pj,Haisch:2016hzu,Domingo:2016yih}. Such mixings can effectively be described through off-diagonal contributions~$\delta m^2_{a\eta_b(n)}$ to the pseudoscalar mass matrices squared. In the case of $a \hspace{0.25mm}$--$\hspace{0.5mm} \eta_b$ mixing, we employ 
\begin{equation} \label{eq:masssquared}
M_{a \eta_b}^2 = \left( 
\begin{array}{cccc} 
m_a^2 - i m_a \Gamma_a & \delta m^2_{a\eta_b(1)} & \ldots  &  \delta m^2_{a\eta_b(6)} \\
 \delta m^2_{a\eta_b(1)} &  m_{\eta_b(1)}^2 - i m_{\eta_b(1)} \Gamma_{\eta_b(1)} & \ldots & 0 \\
 \vdots & 0 & \ddots & 0 \\
\delta m^2_{a\eta_b(6)} & 0 & 0 & m_{\eta_b(6)}^2 - i m_{\eta_b(6)} \Gamma_{\eta_b(6)} 
  \end{array}  \right) \,,
\end{equation}
with 
\begin{equation} \label{eq:offdiagonal}
\begin{split}
\delta m^2_{a\eta_b(n)}  &= \xi_b^{\rm M} \hspace{0.25mm} \sqrt{\frac{3 }{4\pi v^2}  \hspace{0.25mm} m_{\eta_b(n)}^3 }\hspace{0.25mm}   \big |R_{\eta_b(n)}(0) \big  |\,.
\end{split}
\end{equation}
 The masses  and radial wave functions of the $\eta_b (n)$  states are denoted by $m_{\eta_b(n)}$ and $R_{\eta_b(n)}$, respectively.  The latter quantities can be extracted from the $\Upsilon(n)$ leptonic decay widths (see~\cite{Braaten:2000cm} for instance) which are measured rather precisely~\cite{Patrignani:2016xqp}. In the case of $a \hspace{0.25mm}$--$\hspace{0.5mm} \eta_c$ mixing, we only include the first three states in the pseudoscalar mass matrix squared~(\ref{eq:masssquared}) and rely on the potential model calculations of~\cite{Eichten:1995ch} to  determine the radial wave functions $R_{\eta_c(n)}$. The values of the  $\eta_{b} (n)$ and $\eta_{c} (n)$ masses and radial wave functions that are used in our numerical analysis are collected in Table~\ref{tab:metaReta} for convenience. 
 
\begingroup 
\renewcommand{\arraystretch}{1.25}
\setlength\tabcolsep{4pt}
\begin{table}[t!]
\centering
\begin{tabular}{c|c|c|c|c}
&  $m_{\eta_b(n)}$ & $\big |R_{\eta_b(n)} (0) \big |$ & $m_{\eta_c(n)}$ & $\big |R_{\eta_c (n)} (0) \big |$ \\[1mm]   
\hline  $n = 1$ & $9.4$ & $2.71$ &$2.98$ & $0.90$ \\
\hline  $n = 2$ & $10.0$ & $1.92$ & $3.64$ & $0.73$ \\
\hline  $n = 3$ & $10.3$ & $1.66$ &$3.99$ & $0.67$ \\
\hline  $n = 4$ & $10.6$ & $1.43$ & --- & --- \\
\hline  $n = 5$ & $10.85$ & $1.41$ & --- & --- \\
\hline  $n = 6$ & $11.0$ & $0.91$ & --- & --- 
\end{tabular}
\caption{Masses of the $\eta_b (n)$ and $\eta_c (n)$ bound states in units of ${\rm GeV}$ and the corresponding values of the radial wave functions in units of ${\rm GeV}^{3/2}$.}
\label{tab:metaReta}
\end{table}
\endgroup

To be able to determine the eigenvalues and eigenvectors of~(\ref{eq:masssquared}) one also needs to know the total decay widths of the $\eta_{b} (n)$ and~$\eta_{c} (n)$ states. The digluon decay widths of  the $\eta_{b} (n)$ states are given to leading order in $\alpha_s$ by (see~\cite{Drees:1989du} for example)
\begin{align} \label{eq:etagg}
\Gamma \hspace{0.25mm} ( \eta_b (n) \to gg ) & = \frac{\alpha_s^2}{3 m_{\eta_b (n)}^2} \, \big |R_{\eta_b (n)} (0) \big  |^2 \,, 
\end{align}
and an analogous formula holds in the case of the charmonium resonances.  

The partial decay widths~(\ref{eq:etagg}) essentially saturate   $\Gamma_{\eta_b (n)}$ with $n \neq 5,6$. For $\eta_b (5)$ and $\eta_b (6)$, however, also decays to final states involving $\pi$ and $B_{(s)}$ mesons are relevant. In the case of  the decays to pion final states, we employ~\cite{Patrignani:2016xqp}
\begin{align}
\Gamma \hspace{0.25mm} (  \eta_b (5)  & \to \pi \; \text{mesons})  = 1.5 \, {\rm MeV} \,, \\[1mm]
\Gamma \hspace{0.25mm} (  \eta_b (6)  & \to \pi \;  \text{mesons})  = 3  \, {\rm MeV}  \,,
\end{align}
while the $B_{(s)}$ decays are incorporated via the approximate relations~\cite{Baumgart:2012pj}
\begin{align}
& \Gamma \hspace{0.25mm} ( \eta_b (5) \to B +  B_{s} \; \text{mesons})   \simeq 0.9 \hspace{0.5mm} \Gamma \hspace{0.25mm} ( \Upsilon(5)  \to B \; \text{mesons}) +  0.65 \hspace{0.5mm} \Gamma \hspace{0.25mm} ( \Upsilon(5)  \to B_{s} \; \text{mesons}) \,, \\[2mm]
& \hspace{6mm} \Gamma \hspace{0.25mm} ( \eta_b (6) \to B + B_{s} \; \text{mesons})   \simeq  \Gamma \hspace{0.25mm} ( \Upsilon(5)  \to B \; \text{mesons}) +  \Gamma \hspace{0.25mm} ( \Upsilon(5)  \to B_{s} \; \text{mesons}) \,, 
\end{align}
in our numerical analysis. Here~\cite{Patrignani:2016xqp}
\begin{align}
\Gamma \hspace{0.25mm} (  \Upsilon(5)  \to B \; \text{mesons} )  & = 42 \, {\rm MeV} \,, \\[1mm]
\Gamma \hspace{0.25mm} (  \Upsilon(5)  \to B_s \; \text{mesons} )  & = 11 \, {\rm MeV}  \,.
\end{align}

In the case of the charmonium bound states, we use directly $\Gamma_{\eta_{c (1)}} = 31.8 \, {\rm MeV}$ and  $\Gamma_{\eta_{c (2)}} = 11.3 \, {\rm MeV}$~\cite{Patrignani:2016xqp}, while for $ \eta_{c } (3)$ we include besides~(\ref{eq:etagg})  an open-charm contribution. Applying the approach of~\cite{Baumgart:2012pj} to relate the $\eta_c (3)$ decays to those of $\psi (3770)$ results in $\Gamma \hspace{0.25mm} ( \eta_{c } (3) \to D  \; \text{mesons}  ) \simeq 30  \hspace{0.5mm} \Gamma \hspace{0.25mm} ( \psi (3770) \to  D \; \text{mesons}  )$. However, the $\psi (3770) $ lies very close to the open-charm threshold and is thus highly susceptible to strong rescattering effects. Using instead the $\psi (4040)$ properties as input, we obtain the approximate result 
\begin{equation} 
\Gamma \hspace{0.25mm} ( \eta_{c } (3) \to D  \; \text{mesons}  ) \simeq 0.5  \hspace{0.5mm} \Gamma \hspace{0.25mm} ( \psi (4040) \to  D \; \text{mesons}  )  \,,
\end{equation}
where  $\Gamma \hspace{0.25mm} ( \psi (4040) \to   D \; \text{mesons} )  \simeq \Gamma_{\psi (4040)}  = 80 \, {\rm MeV}$~\cite{Patrignani:2016xqp}.

We furthermore emphasise that the branching ratios  $\eta_b (n) \to \mu^+ \mu^-$ are all below the $10^{-10}$ level~\cite{Haisch:2016hzu} and therefore can be safely ignored  in the mixing formalism. The effects of the ditau decays of the bottomonium bound states are negligible as well and so are the dilepton decays of the $\eta_c (n)$ mesons. Effects of $a \hspace{0.25mm}$--$\hspace{0.5mm} \eta_b$ mixing in $h \to aa$ such as for instance $h \to 2 \eta_b (n) \to a a$ are part of ${\rm BR} \hspace{0.25mm} (h \to a a )$ and thus effectively included in our numerical analysis. The same is true for contributions of intermediate $\eta_c (n)$ states to the exotic decay $h \to aa$ of the SM Higgs. 

Above the $b\bar b$ ($c \bar c$) threshold a perturbative description of the production and the decay of the pseudoscalar $a$ breaks down. In this region one can however approximate the $b \bar b$ ($c \bar c$) contributions to the total decay width $\Gamma_a$ through a heuristic model that is inspired by QCD sum rules~\cite{Drees:1989du,Baumgart:2012pj,Haisch:2016hzu} and interpolates to the continuum sufficiently above threshold. The interpolations take the form 
\begin{align}  
{\cal N}^{b}_a & = 1 - \exp \left [ - 8.0  \left ( 1 - \frac{(m_B + m_{B^\ast})^2}{m_a^2} \right )^{2.5 \,} \right ] \,, \label{eq:BastB} \\[2mm]
{\cal N}^{c}_a & = 1 - \exp \left [ - 6.5  \left ( 1 - \frac{(m_D + m_{D^\ast})^2}{m_a^2} \right )^{2.5 \,} \right ] \,, \label{eq:DastD} 
\end{align}
with $m_B = 5.28 \, {\rm GeV}$, $m_{B^\ast} = 5.33\, {\rm GeV}$, $m_D = 1.86 \, {\rm GeV}$ and $m_{D^\ast} = 2.01 \, {\rm GeV}$~\cite{Patrignani:2016xqp}.  In our analysis, the interpolation is achieved by simply multiplying the partonic decay width $\Gamma \hspace{0.25mm} (a \to b \bar b)$  and  $\Gamma \hspace{0.25mm} (a \to c \bar c)$  by the factor ${\cal N}^{b}_{a}$ and ${\cal N}^{c}_{a}$, respectively. 

For $m_a > 2 m_K$ decays into kaons become kinematically allowed. The decay $a \to KK$  however violates CP, and as a result $a$ can in practice only decay into three-body final states such as $KK\pi$. Following~\cite{Dolan:2014ska}, we estimate the hadronic width $\Gamma  \hspace{0.25mm} ( a \to s \bar s \to KK \pi )$ by multiplying $\Gamma  \hspace{0.25mm} ( a \to s \bar s )$ by the suppression factor 
\begin{equation}
{\cal N}^{s}_a  = \frac{16 \pi}{m_a^2} \left ( \frac{m_s^\ast}{m_s} \right )^2 \frac{\rho \left ( m_K, m_K, m_\pi, m_a \right ) }{\beta_{s/a}} \,,
\end{equation}
with $m_s^\ast = 450 \, {\rm MeV}$~\cite{McKeen:2008gd}, $m_K = 439 \, {\rm MeV}$ and $m_ \pi = 140 \, {\rm MeV}$~\cite{Patrignani:2016xqp}. Here  $\rho \left ( m_1, m_2, m_3, m_4 \right )$ denotes the phase space for isotropic three-body decays. It can be written as  
\begin{equation}
\begin{split}
\rho \left ( m_1, m_2, m_3, m_4 \right ) & = \frac{1}{(4 \pi)^3} \int_{m_1}^{\frac{m_1^2 + m_4^2 -(m_2+m_3)^2}{2m_4}} \! d E_1 \, 2 \hspace{0.25mm} \sqrt{E_1^2 - m_1^2} \\[2mm] 
& \hspace{1.75cm} \times \lambda \left ( m_1^2 + m_4^2 - 2 E_1 m_1, m_2^2, m_3^2 \right )  \,,
\end{split}
\end{equation}
with $\lambda \left (x, y, z \right)$ defined in~(\ref{eq:lambdaxyz}). 

\end{appendix}


%

\end{document}